\documentclass[english,aps,prper,reprint,showpacs,titlepage,longbibliography]{revtex4}   

\usepackage[T1]{fontenc}	
\usepackage[latin9]{inputenc}	
\usepackage{geometry}		
\usepackage{graphicx}
\usepackage[above,below]{placeins}	
\usepackage{times}

\usepackage{hyperref}  
\hypersetup{colorlinks=true,urlcolor=blue,citecolor=blue,linkcolor=blue}   
\usepackage{enumitem}          
\setlist{nosep}                 

\usepackage{censor}

\begin{document}

\begin{titlepage}

  \title{Student evaluation of more or better experimental data in classical and quantum mechanics}

  \author{Courtney L.~White}
  \affiliation{Physics Department, California State University, Fullerton, 800 N State College Blvd, Fullerton, CA, 92831} 
  \author{Emily M.~Stump}
  \author{N.~G.~Holmes}
  \affiliation{Laboratory of Atomic and Solid State Physics, Cornell University, 245 East Avenue, Ithaca, NY, 14853} 
 
  \author{Gina Passante}
  \affiliation{Physics Department, California State University, Fullerton, 800 N State College Blvd, Fullerton, CA, 92831~} 


  \begin{abstract}
      Prior research has shown that physics students often think about experimental procedures and data analysis very differently from experts. One key framework for analyzing student thinking has found that student thinking is more point-like, putting emphasis on the results of a single experimental trial, whereas set-like thinking relies on the results of many trials. Recent work, however, has found that students rarely fall into one of these two extremes, which may be a limitation of how student thinking is evaluated. Measurements of student thinking have focused on probing students' procedural knowledge by asking them, for example, what steps they might take next in an experiment. Two common refrains are to collect more data, or to improve the experiment and collect better data. In both of these cases, the underlying reasons behind student responses could be based in point-like or set-like thinking. In this study we use individual student interviews to investigate how advanced physics students believe the collection of more and better data will affect the results of a classical and a quantum mechanical experiment. The results inform future frameworks and assessments for characterizing students thinking between the extremes of point and set reasoning in both classical and quantum regimes.

   \clearpage
  \end{abstract}

  \maketitle
\end{titlepage}

\section{Introduction}

Experiments are of critical importance in physics.  In physics instruction, laboratory exercises comprise a large component of our courses at both the high school and university level. While different lab courses might have a different focus, all of them require students to collect data, interpret that data, and then draw conclusions.

%
Previously, student thinking about measurement and uncertainty was described using two paradigms: the point and set paradigms~\cite{buffler2001development}. In the point paradigm, students make decisions and draw conclusions based on the result of a single experimental trial. In contrast, in the set paradigm, students use a set of measurements in order to make claims about the results. People have identified limitations with these two paradigms, in particular that a significant proportion of students use mixed reasoning: reasoning point-like on some questions and set-like on others~\cite[e.g.][]{pollard2018impact, Lubben2001}. 

Students' point-like or set-like reasoning was based on evaluations of students' procedural knowledge: suggestions for what should be done (procedurally) in an investigation~\cite{buffler2001development}. For example, students are asked what value(s) they would report from collected data, to compare measurements, or what one should do next in an investigation. While some procedural statements are clearly point-like or set-like, some statements might be either. For example, two common suggestions for next steps are to take more data or perform the experiment with better equipment~\cite{Sere1993, HolmesBFY2015}. Suggesting to take more data can be both point-like (students want to take more data so the experimenter improves their technique so a future data point will be closer to the true value) or set-like (more data will produce a bigger set of data which makes for a better estimate of the true value). Improving the equipment, such as by having an expert in a research lab perform the experiment, may also be point-like (such that the expert can measure perfectly in the lab and reduce the uncertainty to zero~\cite{Leach1998, Kung2006}) or set-like (such that the expert can never reduce the uncertainty to zero, but can decrease it). Thus, it is important to not only understand what procedures students think should be carried out, but also \emph{why} they make these decisions.

Our goal is to better understand what impact students think each of these options (more data and better data) would have on a data set, offering a tangential perspective from the procedural knowledge previously studied. Furthermore, we wanted to explore this reasoning in multiple contexts. Of note, most of the research on these topics has focused on students at the introductory physics levels and classical mechanics scenarios. Previous research, however, has found that students think deterministically about both classical measurement~\cite{buffler2001development} and about quantum mechanics~\cite{Kalkanis2003}. How might students' understandings of the effect of more data or expert data differ between classical and quantum mechanics measurements?





The research questions for this work are: how do students believe that obtaining (a) more data or (b) better data will affect a histogram of many repeated measurements?  We are investigating these questions in both a classical context, such as an experiment students might come across in an introductory physics lab, and in a quantum mechanical context. The overarching goal of this work is to further understand how students think about measurement and uncertainty across physics contexts through a perspective than goes beyond procedural knowledge. 


\section{Methods}

\subsection{Research Context}
We performed individual interviews with 19 advanced physics students (juniors, seniors, and masters-level students) split across two universities in the United States. One institution is a private, selective, research-intensive institution in the northeast and the second is a public, Hispanic-serving, teaching-focused institution in the southwest. Interviews were video recorded for later transcription and analysis.  Participants were compensated for their participation. Note that a previous analysis of these data can be found in 

The interview questions analyzed here were split into two parts: a ball drop experiment and a single slit experiment.  Both of these experiments were hypothetical and the procedures for each were explained to the interview participants. The ball drop experiment was taken from the Physical Measurement Questionnaire~\cite{buffler2001development}, where a ball is placed at the top of a ramp that is sitting on a table.  The ball rolls down the ramp and flies through the air before hitting the ground. The location the ball lands away from the table is recorded.  In the single slit experiment, individual photons are incident on a small single slit.  The photons then travel to a screen and the position of each photon is recorded.  

\begin{figure}
  \includegraphics[scale=0.55]{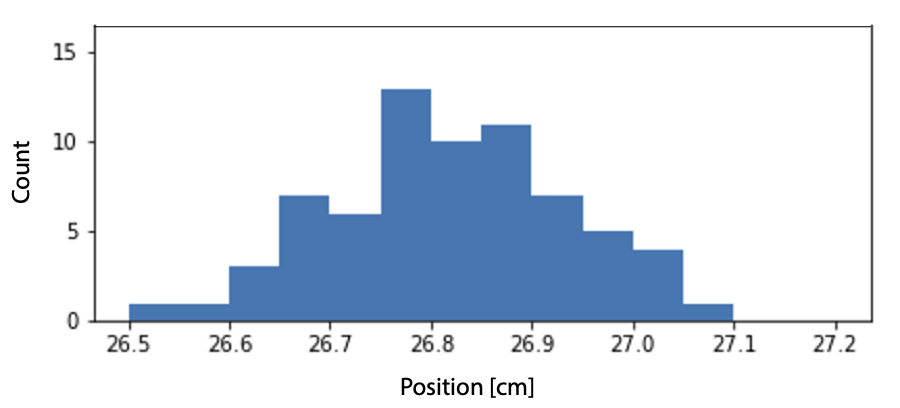}
  \caption{Histogram that interview participants were shown for both the ball drop experiment and the single slit experiment (with slightly different labels along the horizontal axis so that the center of the histogram was around 2 cm instead of 26.8 cm as shown).\label{fig:histogram}}
\end{figure}

During the interview, it was explained that these experiments were performed by students in a lab course and each student collected a single data point.  These data were then combined into a histogram showing the results (see Fig.~\ref{fig:histogram}). The same histogram was shown for both the ball drop and the single slit experiment with the exception of the labels along the horizontal axis.  

The interviews were structured with the same questions asked in both the ball drop and the single slit contexts. For both experiments, we asked participants specific questions about the histogram given and what they think about data and results. In the experiments, the same questions were used to ensure comparisons could be drawn concluding the interviews.  In this work we will discuss student responses to two questions that were designed to probe how the histogram would change if more data were added or if better data were collected instead. The `more data' element was addressed by the question: \textit{"Now a hundred students come to class late and do their measurements. What do you expect the distribution to look like with the additional data?}  This question is asked after the participants had answered questions about the histogram such as how they would characterize it in a few meaningful numbers and what are some of the reasons for the spread in the data.  The `better data' element was addressed by asking: \textit{"If an expert scientist at a national laboratory would take measurements using high precision equipment, what would the resulting histogram look like?"} This was the last question asked in each context.

\subsection{Data Analysis}
The coding scheme was developed to analyze how participants might expect the histogram to change when more data points were added or when better data were taken instead. Responses were coded using four categories: Larger Distribution (LD), More Narrow (MN), More Accurate (MA), and No Change (NC). These codes were determined after preliminary reading of interview transcripts in order to capture all responses. 

The LD (larger distribution) and MN (more narrow) codes, refer to the precision of the experiment.  As one might expect, the LD and MN codes are assigned when a student indicates using words or drawings that the histogram will increase in width or decrease in width, respectively.  
The MA (more accurate) code is reserved for when students discuss the position of the peak of the histogram, or the `true value' or `expected value' explicitly. 
A NC (no change) code is for when students explicitly state that there will be no change to the spread of the histogram or that it will look similar to the original.  For completeness, there could be a less accurate code, but this was never mentioned by students, so it is not included.  

These codes are not orthogonal, and therefore it was possible for responses to be assigned multiple codes.  While this was not common, it did occur occasionally with the MA and MN codes. On the other hand, some students did not provide explanations that could clearly be coded into any of these categories, and were therefore not coded.  

In the original coding scheme, there was a fifth code for students that indicated the histogram would become a delta function. The researchers believed that students might be likely to indicate that an expert might get the `exact' answer, resulting in a delta function distribution, as previously observed~\cite{Leach1998}.  However, none of our interview responses indicated that the histogram would lose all uncertainty.

All transcripts 
were independently coded by the first two authors.  After initial agreement of 81\%, all disagreements were discussed until 100\% agreement was reached.

\section{Results}

We separate our discussion of the results into two sections, one for the ball drop (or classical) experiment and one for the single slit (or quantum) experiment.  In each context we discuss student responses to the more and better data questions separately before comparing them.  In the Discussion (Section~\ref{sec:Discussion}) we compare classical and quantum results directly.

\subsection{Classical: Ball Drop Experiment}
\begin{figure}
        \includegraphics[width=0.42\textwidth]{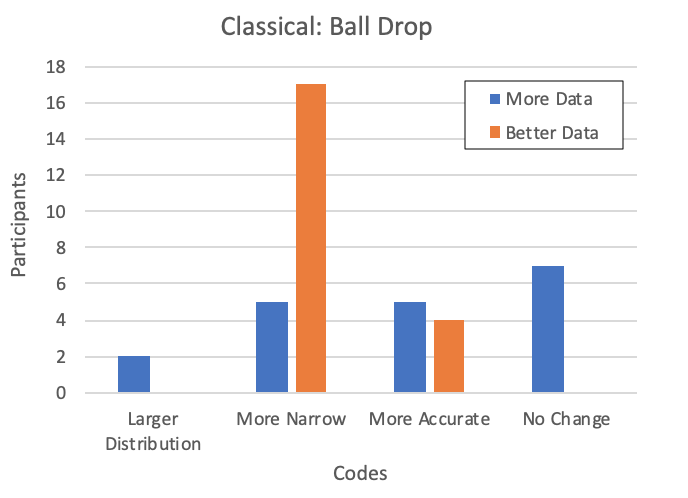}
        \caption{Number of responses coded in each category for the ball drop experiment.  Note that some responses may have been coded in multiple categories or not given a code.  ($N = 19$) \label{fig:classical}}
        \end{figure}

Results for both more and better data in the ball drop experiment can be seen in Fig.~\ref{fig:classical}.
        
\subsubsection{More Data}
When asked what the histogram would look like if 100 more students took data in the ball drop experiment, many participants stated that the histogram would have no change to the spread: \textit{"Yeah. But the standard deviation and the mean should remain the same."} 

There were also a number of participants who described the more data histogram as getting more accurate and/or more narrow. The more narrow code was assigned when students described with words or drawing that the histogram would decrease in width: \textit{"I mean, so it's going to have, I think really similar shape, but everything is going to be higher. Um, it might narrow a bit."}  Other students explicitly stated that the data would become more accurate: \textit{"I mean, it gets you closer to the real value."}  There were two students who indicated that the histogram would become both more narrow and more accurate:  \textit{"Uh, so you're getting fewer outliers and more people landing precisely on, uh, the calculated expected value in an ideal situation."}

    
While LD was the least common response, there were a couple of participants who expected a larger distribution once additional data were added: \textit{"...but also there would probably be like a few additional outliers because there are more measurements being made."} 

\subsubsection{Better Data}
When asked what the histogram would look like if an expert scientist were to perform the experiment, many more responses were coded as more narrow than when students were asked about more data. Most participants justified this by stating that the expert would have better equipment: \textit{"...because his just like, uh, equipment, is more precise so you can sort of, uh, more, more, um more accurately control the parameters and make sure that each experiment is very similar..."} Participants also mentioned repeatability as a reason for a more narrow distribution: \textit{"Uh, I'd expect it to be much narrower, uh, because they'll just have this experimental set up that's more repeatable."} As previously observed~\cite{Leach1998}, most of the interview participants think the expert can do a better job than 100 students, but surprisingly did not think that the expert would get the exact value (i.e., a delta function). 
    
In contrast, relatively few responses indicated that the expert scientist's experiment would lead to a more accurate histogram. Those participants who did mention that a more accurate histogram would result explained that the expert is less likely to make mistakes: \textit{"So, if this is the theoretical spot right here, then in a professional lab where they have better techniques and it should be more kind of localized."} No responses were coded as expecting a larger distribution nor that there would be no change to the histogram.


\subsubsection{Comparison: More vs.~Better}         
Using Fig.~\ref{fig:classical} as a guide, we can see that participants respond very differently to more and better data affecting the distribution.  In response to more data, we see that participant responses are almost evenly distributed between the more narrow, more accurate, and no change codes.  We even see some participants indicating that the distribution will become wider.  In contrast, better data was coded to produce a more narrow histogram by all but two participants.  Notably, not a single participant indicated that the expert would only obtain a single value, which would result in a delta function.  We also found that relatively few participants indicated that the expert scientist would produce a more accurate histogram, compared to more narrow.  This might be due to the nature of the question, which might not have cued students to think about the placement of the peak but rather the spread only.


\subsection{Quantum: Single Slit Experiment}
\begin{figure}
        \includegraphics[width=0.42\textwidth]{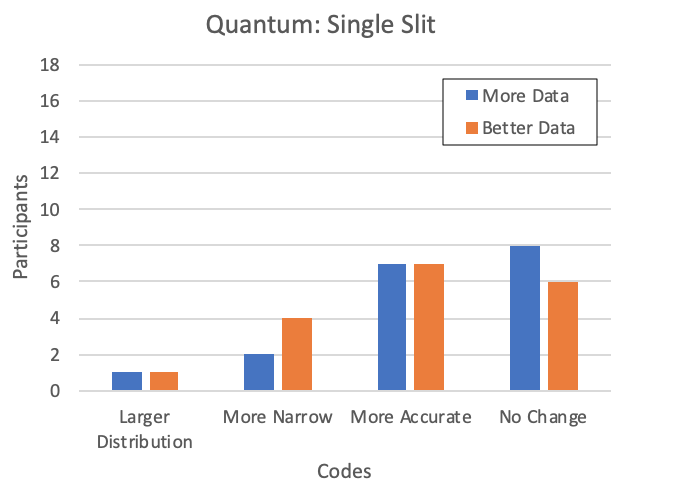}
        \caption{Number of responses coded in each category for the single slit experiment.  Note that some responses may have been coded in multiple categories or not given a code.  ($N = 19$)  \label{fig:quantum}}
        \end{figure}

In this section we analyze participant responses to the same set of questions, but this time in the context of the single slit experiment.  The results are shown in Fig.~\ref{fig:quantum}.     
     
\subsubsection{More Data}
There were two common types of responses when students were asked about more data: no change and more accurate.   Explanations for why students believe there will be no change tended to be fairly vague, as shown here: \textit{"Because here I would expect it to look more or less the same as if, um, if many students, uh, measure it."} 
Responses coded as more accurate indicated that 
the true value could be achieved at some point: \textit{"I mean the more point you have, the better your curve is going to look in. The closer [to the true value] your data data will be."} This is consistent with the idea that the more data you have, the better the results will be.

In the single slit experiment, very few responses indicated that the distribution would become more narrow: \textit{"...instead of the curve like that, it'll become more steep on the sides and have a larger peak."} Only a single response indicated that more data would lead to a larger distribution.
    
\subsubsection{Better Data}

When asked what the histogram might look like for an expert scientist, some participants indicated that histogram would be more accurate:  
\textit{"Um, I think it'll give you a better picture if that makes sense. Something closer to like what it should be."} Another participant explained: \textit{"Uh, just more, more, more precise, more, more numbers where the data shows it should be in terms of theoretical."} (Note that although this participant used the word `precise', it was clear from context and visualizations that they meant more accurate.  It is not unusual for students to conflate these terms~\cite{Sere1993}.) Participants' reasoning for why the data would be more accurate was varied and there was no one common explanation.
    
Many participants also indicated that there would be no change in the histogram when an expert scientist conducts the experiment:\textit{"...probability spread is just in the nature of the experiment and not necessarily in any error of the setup of the system. Uh, I expect a scientist to see the same kind of distribution that the students are already seeing."} Many of the explanations explicitly indicated that it is not possible to eliminate the uncertainty in a quantum experiment.
    
A few participants indicated that the expert scientist would get either a more narrow or a larger distribution. Responses for the histogram to become more narrow suggested that the expert could eliminate some amount of error: \textit{"It might be slightly narrower since um, there might be I guess some measurement error."} While other participants stated that an expert scientist would pick up more background with the most precise equipment: \textit{"
\ldots it's not necessarily like steeper, it's just bigger."} This particular response was accompanied by a drawing that clearly depicted a larger distribution. The single response indicating there would be a larger distribution for both more and better data were given by two different participants.

\subsubsection{Comparison: More vs. Better}
In the context of the single slit experiment, student responses were remarkably similar for both more and better data.  In describing why the results might become more accurate, participants tended to provide vague reasons such as more data gives better results and in the case of the expert scientist, they sometimes indicated the better equipment would be more accurate. 
A similar trend occurred for participants that said there would be no change to the histogram.  They indicated that more data would just result in more of the same, however to justify why an expert would get the same results, they tended to lean on the ideas of quantum uncertainty.


\section{Discussion}
\label{sec:Discussion}

The point and set paradigms for physical measurements as defined by Buffler and colleagues~\cite{buffler2001development} have been one of the main frameworks for analyzing student thinking about laboratory measurements. However, student thinking may not clearly fall into one of these two extremes~\cite{pollard2018impact, Lubben2001}, suggesting that new ways of probing student thinking about measurement are necessary. 
In our study, we asked students specifically how collecting more data and better data would affect their findings. 

Evidence of student thinking about measurement under the point paradigm would be evident by claims that experts should get the `right' answer, resulting in a narrow peak that approaches a delta function 
at the location of the `true value'.  Our findings show that while almost all participants in our study indicated that better data would result in a narrower histogram in the classical context, not a single participant indicated that a delta function would result.   

Evidence of student thinking about measurement using the set paradigm might have been evident by 
students thinking that more data would allow for more statistical power, as it would reduce the standard error. We found some evidence of possible conflation between a reduction of the standard error and the standard deviation, as it was not uncommon for participants to indicate that the histogram would get more narrow (precise) 
as a result of more data.  This is in keeping with the literature that suggests students ``rarely understand what these summary statistics represent''~\cite[][p.385]{Garfield2007}. It further clarifies that students' responses to procedural questions (such as through suggestions to take more data~\cite{Sere1993}) may reflect habitual responses~\cite{Bakker2011}. The questions analyzed in this paper revealed nuances in student understanding of measurement previously missed.


Additionally, our data provide new insights into possible manifestations of point- and set-like thinking in the context of quantum mechanical experiments.  The biggest difference between the classical and quantum contexts were the number of students who felt that better data would result in a more narrow histogram. This may be due to the fact that students appropriately expected a single slit experiment to result in a diffraction pattern.  However, this does not necessarily mean that students are thinking set-like in quantum mechanics, as it is still possible to think deterministically on the individual photon level and obtain a diffraction pattern. Claims about point- and set-like thinking in quantum mechanics requires further study.


The analyses here provide new evidence for probing student thinking about measurement uncertainty in multiple contexts. While we interviewed a diverse sample of students from very different institutions, the results here would benefit from additional testing with a larger and broader population of students. In future work, we will use these results to develop new assessments and refine paradigms for student thinking about measurement uncertainty in classical and quantum mechanics, and possibly other contexts.




\acknowledgments{
The authors would like to thank the members of their respective research groups for comments on this manuscript.  
This work has been supported in part by the NSF under Grants No.~DUE-1808945 and No.~DUE-1809183.
} \newpage


\bibliography{refs} 


\end{document}